\documentclass[english]{article}
\usepackage[utf8]{inputenc}
\usepackage[T1]{fontenc}
\usepackage{amsmath}
\usepackage{graphicx}
\usepackage{fancyhdr}
\usepackage{float}
\usepackage{subcaption}
\usepackage{adjustbox}
\usepackage{multirow}
\usepackage{hyperref}
\newcommand{\scidatalogo}{\includegraphics[height=36pt]{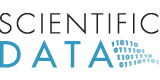}}
\newcommand{\overleaflogo}{\includegraphics[height=36pt]{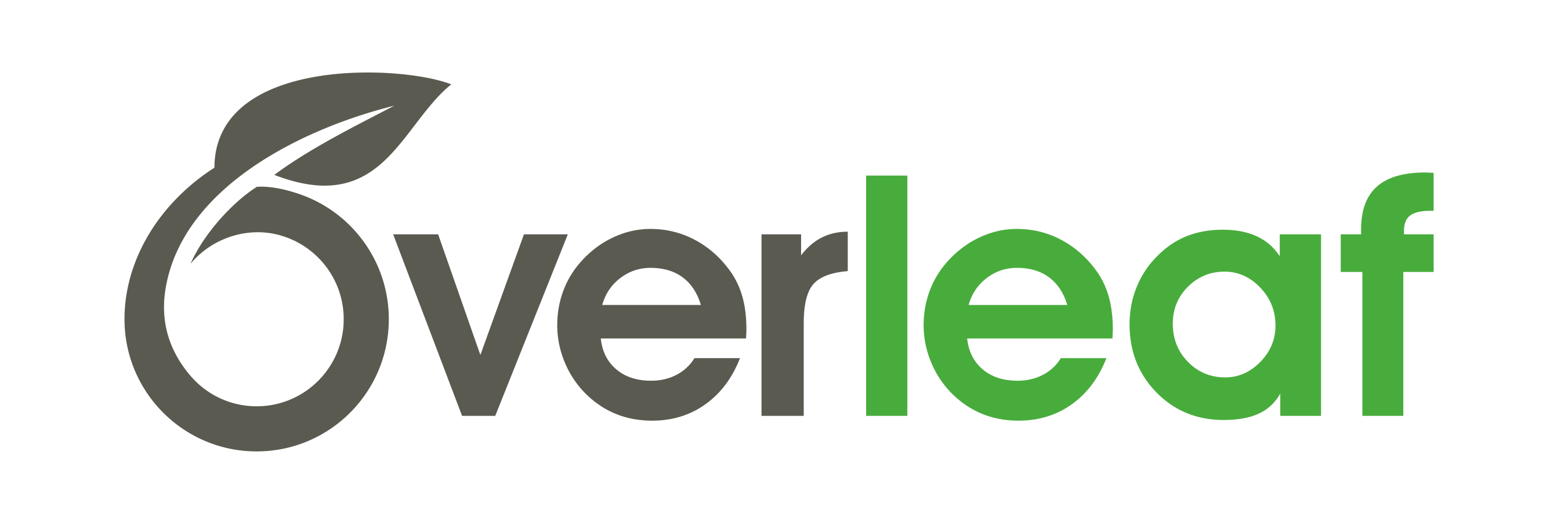}}
\begin{document}

\title{Explorative pedestrian mobility GPS data from a citizen science experiment in a neighbourhood}

\author{Ferran Larroya\textsuperscript{1,2}, Roger Paez\textsuperscript{3}, Manuela   Valtchanova\textsuperscript{3}, Josep Perell\'o\textsuperscript{1,2,{*}}}

\maketitle
\thispagestyle{fancy}

1. OpenSystems, Departament de F\'isica de la Mat\`eria Condensada, Universitat de Barcelona, Martí i Franquès, 1, 08028 Barcelona, Catalonia, Spain 

2. Universitat de Barcelona Institute of Complex Systems, Universitat de Barcelona, Barcelona, Catalonia, Spain

3. ELISAVA Barcelona School of Design and Engineering, Universitat de Vic - Universitat Central de Catalunya, La Rambla, 30-32, 08002 Barcelona, Catalonia, Spain

{*}corresponding author(s):
Ferran Larroya (ferran.larroya@ub.edu) and Josep Perell\'o (josep.perello@ub.edu)

\begin{abstract}
Pedestrian GPS data are key to a better understanding of micro-mobility and micro-behaviour within a neighbourhood. These data can bring new insights into walkability and livability in the context of urban sustainability. However, pedestrian open data are scarce and often lack a context for their transformation into actionable knowledge in a neighbourhood. Citizen science and public involvement practices are powerful instruments for obtaining these data and take a community-centred placemaking approach. The study shares some 3\,000 GPS recordings corresponding to 19 unique trajectories made and recorded by groups of participants from three distinct communities in a relatively small neighbourhood. The groups explored the neighbourhood through a number of tasks and chose different places to stop and perform various social and festive activities. The study shares not only raw data but also processed records with specific filtering and processing to facilitate and accelerate data usage. Citizen science practices and the data-collection protocols involved are reported in order to offer a complete perspective of the research undertaken jointly with an assessment of how community-centred placemaking and operative mapping are incorporated into local urban transformation actions.
\end{abstract}

\section*{Background \& Summary}

Mobility and public spaces are both identified as key issues in the attempts to achieve more sustainable, walkable and livable cities \cite{Ravazzoli2017}. Understanding how cities affect the behaviour of their inhabitants and how urban public spaces can be reshaped by this behaviour are important challenges to be considered in the development and regeneration of public spaces \cite{Carmona2018}. The analysis of available digital traces \cite{jiang} can contribute to this effort as part of the interdisciplinary approaches currently applied within the field of urban science \cite{Batty2013,Bibri2021,Bettencour2021}. Some of the researchers in this field are taking a complex systems science approach and human mobility data represent one important data source \cite{Batty2013,Bettencour2021,Bai2016}.

When studying urban mobility, digital traces have shown their potential for investigating long distance movements (e.g., in public transportation \cite{clemente}) but there is still a lack of high resolution micro-mobility data to study and characterize the local micro-behaviours in people's neighbourhoods \cite{hunter,emoro}. Many research questions may be related to micro-mobility. Among them, it would be of interest to know how people use the public spaces for walking or which places they choose for performing specific activities. The inferred walkability taking GPS data in its finest resolution overlaps strongly with the analysis of livability in general, and with lively and sociable urban environments in particular \cite{Tobin2022}. In this context, the street-space approaches of urban placemaking \cite{Thomas2016} and operative mapping \cite{Paez2019} can promote and encourage a more open and collaborative use of high-resolution GPS data, create reflection and discussion among citizens and city makers, and motivate public space transformations with new evidence \cite{Valtchanova2023, paez2024}.

Participatory research approaches such as citizen science \cite{citizenscience} represent a valid option for providing high resolution datasets in small-scale urban contexts \cite{citizenscience2, citizenscience3} and contribute to a community-centred placemaking in more livable cities \cite{Palmer2024}. Citizen science practices involve the active participation of the general public in scientific research tasks, which may include the formulation of scientific questions, the elaboration of the experimental protocol, and the collection and subsequent interpretation of data \cite{citizenscience}. Citizen science data are already contributing to the monitoring of sustainable goals in a bottom-up approach \cite{Ballerini2021} but social injustices emanating from the data processes in citizen science need to be considered in order to avoid exploitative and exclusionary implications \cite{Christine2021}. To effectively contribute to community urban planning \cite{Sandercock2023}, citizen science thus should incorporate diversity of participants, consider social and environmental justice, and co-produce new knowledge responding to people's concerns in an actionable manner \cite{Albert2021,Bonhoure2023}.

\begin{figure}
\centering
\includegraphics[width=0.6\textwidth]{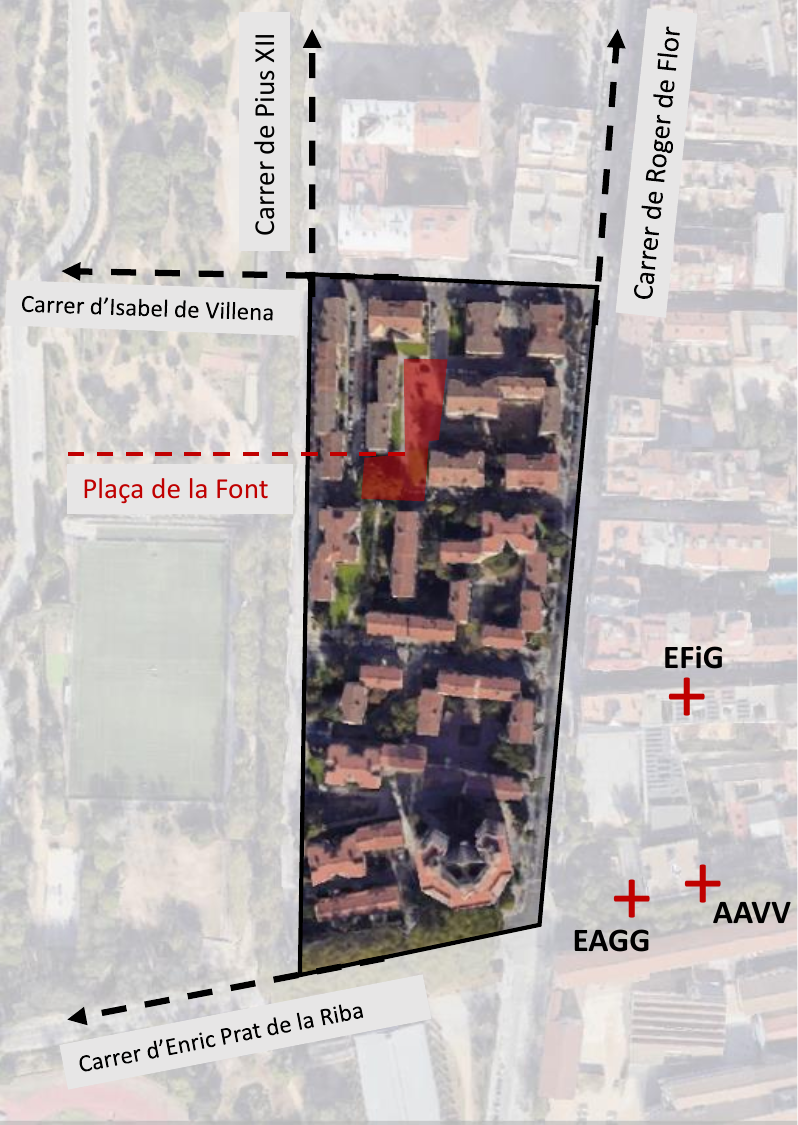}
\caption{{\bf Map of ``Barri Primer de Maig''.} The neighbourhood ``Primer de Maig'' in the city of Granollers (Catalonia, Spain) is limited by the streets: ``Carrer d'Enric Prat de la Riba'', ``Carrer de Roger de Flor'', ``Carrer de Pius XII'' and ``Carrer d'Isabel de Villena''. The red crosses represent the location of the three sites related to the communities involved: the school ``Escola Ferrer i Guàrdia (EFiG)'', the neighbourhood association site ``Associació de Veïns Sota el Cami Ral (AAVV)'', and the space for older people ``Espai Actiu de la Gent Gran (EAGG)''. The meeting point was located at the ``Pla\c ca de la Font'' and this square is also displayed in red. The neighbourhood has an area of approximately 3 ha (300$\times$100 metres) and the distance between the three communities' locations and the neighbourhood is less than 70 metres. The satellite image was retrieved from the Spanish National Geographic Institute (\url{https://www.ign.es/web/ign/portal}) with data from the European Copernicus satellites.}
\label{fig:map_granollers}
\end{figure}

The present study focuses on the community action to explore temporary uses of public spaces in the ``Primer de Maig'' neighbourhood of the city of Granollers (Catalonia, Spain). The city belongs to the Barcelona metropolitan region and  has 61\,983 inhabitants, covers 1\,487 ha and 4\,160 inhabitants per ${\rm km}^{2}$. The ``Primer de Maig'' neighbourhood is one of the 35 censal districts of the city, with a surface area of about 3 ha and 45\,200 inhabitants per ${\rm km}^{2}$ (1\,356 residents). Figure \ref{fig:map_granollers} shows an aerial map of the neighbourhood. It is the district with the lowest net annual income per person (9\,329 EUR) and per dwelling (25\,026 EUR) (\url{https://www.ine.es/jaxiT3/Tabla.htm?t=30896}, in Spanish). This small neighbourhood was one of the first areas of peripheral homes in the city of Granollers. The buildings provided a rapid solution to the housing problem that arose in the middle of the 20th century due to the large-scale migration waves (\url{https://www.granollers.cat/can-jonch/grup-primer-maig}, in Catalan). Until that time, this area was mostly devoted to industry and agriculture. The neighbourhood is made up of 43 buildings, with a total of more than 550 homes. Its urban nature is based on the construction of typologies of repetitive housing blocks with a strong presence of blind party walls and surrounded by medium-small empty spaces. Public space is generous in terms of quantity but poor in terms of quality. These characteristics mark the idiosyncrasy of the neighbourhood and its possibilities of transformation.

\begin{figure}
\centering
\includegraphics[width=1\textwidth]{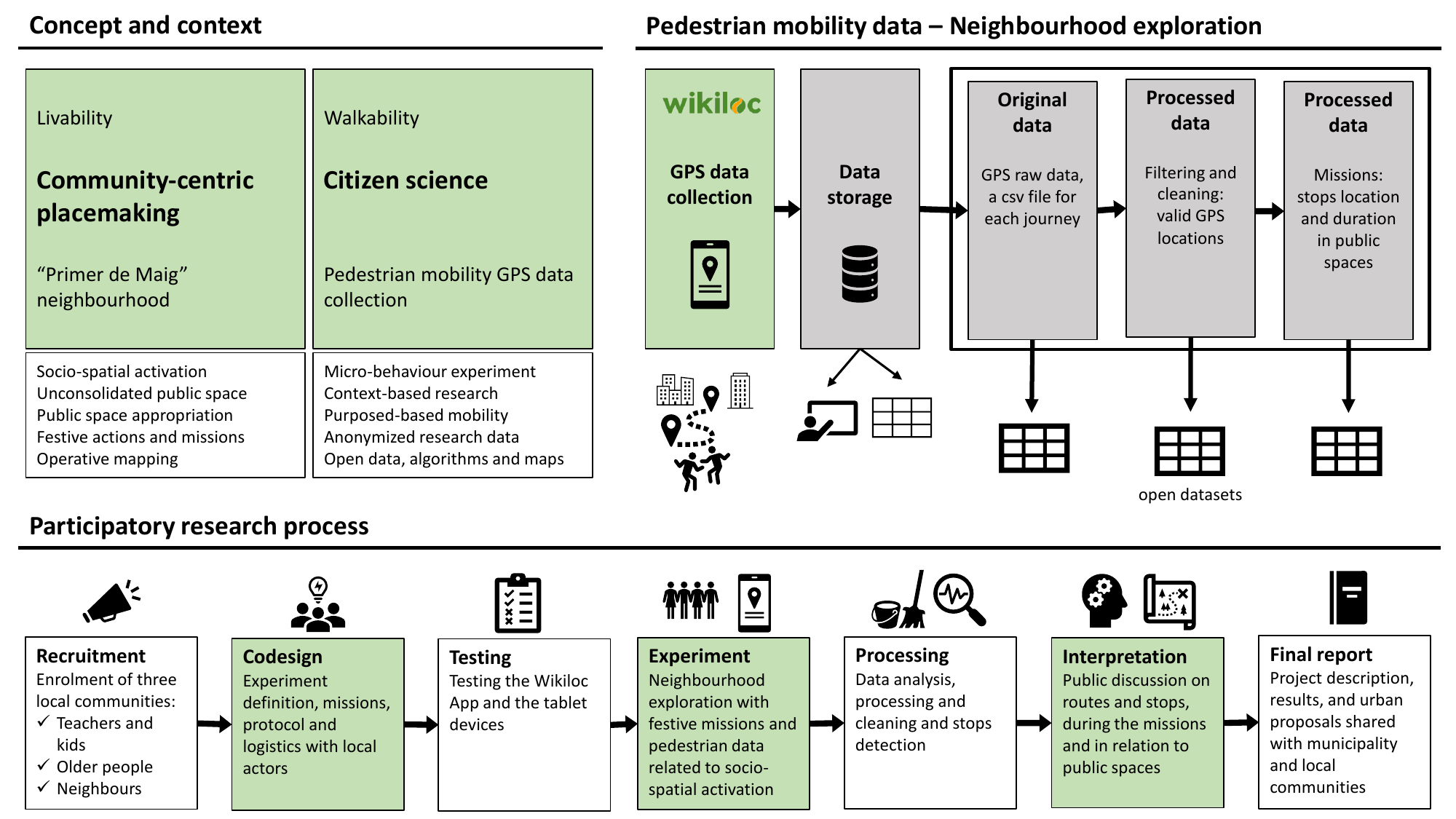}
\caption{{\bf Schematic description of the citizen science experiment.} The data files available are reported jointly with the participatory process underpinning the whole research. The direct participation of citizens is highlighted in green and the data sets being created in grey.}
\label{fig:diagram}
\end{figure}

Design for City Making Research Lab (Elisava Research, UVic-UCC) invited OpenSystems group (Complexity Lab Barcelona, UBICS) to join the experiment in Granollers  (see Figure \ref{fig:diagram} for details). The broad aim was to enrich and widen urban planners' visions by providing concrete evidence through GPS data to support the redesign of public space and the regeneration of ``Primer de Maig'' neighbourhood \cite{Elisava2022}. The collection of GPS explorative trajectories was based on citizen science practices and on previous experiences in running public and participatory experiments \cite{Sagarra2016,Perello2022,Perello2023}. In 2012, a hundred citizens were engaged in an experiment in a public park (Parc de la Ciutadella, 4 ha, Barcelona) during a Science Festival \cite{Gutierrez2016}. Another experiment (2017) involved 10 public schools from the Barcelona metropolitan area and more than 400 participants, including students (3rd and 4th grade of high school) and teachers \cite{Larroya2023}. In both cases, we collected data on purpose-based pedestrian mobility and built our own mobile App to obtain the GPS data from participants' mobile phones. However, periodic updates and maintenance of the App represented important logistic challenges. 

For the experiment in the ``Primer de Maig'' neighbourhood, we used apps already existing on the market to track routes in different outdoor sports activities (Wikiloc App, \url{https://www.wikiloc.com}). The App being used was tested to confirm that it recorded high quality GPS data at a high resolution of about 4-5 seconds maximum in movement (when the App does not detect any movement, the data collection is paused). The Wikiloc App allows the registration of users with a Google account, and then gives access to the recorded trajectories and downloads the data. As is known, sharing GPS data entails some additional privacy risks and efforts must be made and to avoid possible identification of participants based on the possibility of locating their homes. In these cases a geo-masking technique ($k$-anonymity) has to be applied \cite{Larroya2023,kanonymity}. The experiment in  the ``Primer de Maig'' neighbourhood avoided this risk with a privacy-by-design approach as the participants explored the neighbourhood in groups of at least three, and started and finished the experiment at pre-established public locations that served as meeting points. By using rented devices with the research staff's Google accounts activated, we also prevented participants from providing any private personal information. The protocol guaranteed the total anonymity of the participants, who also read and signed an informed consent. For all these reasons, we are able to share both raw and processed GPS trajectories for all journeys.

The participants represented different socio-demographic profiles which were all considered strategic agents in the neighbourhood. The groups were invited to explore the neighbourhood and find places for different activities in the form of tasks. As part of the ``Let's Celebrate'' ({\it Fem la Nostra Festa}, in Catalan) event, the tasks were planned to obtain socio-spatial activation of unconsolidated public spaces. They aimed to connect spatiality (e.g., physical, material and infrastructural conditions of the neighbourhood) and sociality (e.g., relational, cultural and performative components) \cite{Sennett2018}. The festive format of the activity aimed to reinforce an active attitude to connecting urban space and participants' desires. The GPS data collection was complemented with additional activities to share and discuss the aggregate GPS data from all groups. ``Let's Celebrate'' triggered proactive and creative urban attitudes to discovering opportunities for socialization, to identifying the coexistence between different appropriations of public space, and to promoting sustained social development based on complexity and diversity.

Figure \ref{fig:diagram} shows the schematic description of the process. Here we share and report unique micro-mobility data from 72 participants (19 groups) with different socio-demographic traits. Participants acted as explorers of the neighbourhood, carrying around an exploration kit with different objects to perform specific festive tasks in places of their choice  (see Table \ref{tab:missions}). We share the original datasets and processed records with specific filtering and processing procedures for scientific research in order to facilitate data re-usability \cite{granollersdata}. These micro-mobility data represent a powerful source of data in the field of exploratory mobility. From an academic scientific perspective, data can indeed enrich and contribute to discussions about the most suitable models for exploratory pedestrian mobility \cite{barbosa} (e.g., through stochastic or random walk models). Mobility can also be characterized by velocities (see Methods) or other statistical variables like turning angle or tortuosity \cite{barbosa}. Finally, from the perspective of data science, clustering techniques can be applied, for example to find high stop density regions, or to study the direction of movement flows in the busiest areas. This would help architects and urban planners to plan actions and interventions in public space in order to enhance livability. 

\section*{Methods}

The Universitat de Barcelona Ethics Committee (IRB00003099) approved this mobility experiment. All participants read and signed the informed consent form and parental/legal guardian consent if appropriate. No privacy issues were observed to be in conflict with the public release of the underlying processed data.

\subsection*{Co-designing the experiment with local communities}

Between October 2021 and January 2022, the Design for City Making Research Lab and the OpenSystems group joined forces to design an urban mobility experiment in which the analysis of the data obtained could help to understand how the public space of the neighbourhood ``Primer de Maig'' in Granollers could be redesigned. The researchers decided to develop a set of public experiments \cite{Sagarra2016,Perello2022,Perello2023} in groups of people from local communities, where the participants had to explore the public space while choosing specific sites to perform  actions of a festive nature, and carrying a festive exploration kit with various objects to perform the actions, see Table \ref{tab:missions}). The trajectories (GPS records) followed by the participants were then registered through the Wikiloc App using tablets. The experiments were planned as part of the neighbourhood festivities (April 29 and 30 and May 1, 2022). 

\begin{table}[t]
\centering
\caption{\label{tab:missions} \textbf{List of tasks to complete during the experiment.} The groups had to choose a specific location in the neighbourhood to perform each of the six tasks from the list. A kit comprising items for the festivities was provided. The tablets that each groups was carrying recorded GPS data, and filmed a video of the groups completing each action. This latter activity was a part of the data validation process.}
\begin{adjustbox}{width=\textwidth}
\begin{tabular}{lll}
\hline \hline
Task & Code & Instructions \\ \hline
Dance & B & Explore and choose a place to dance using the speaker via Bluetooth \\
Play & J & Explore and choose a place to play with the ball\\
Chat & X & Explore and choose a place to sit and share personal neighbourhood stories\\
Lunch & M & Explore and choose a place to lay out the carpet and sit down to eat/snack \\
Celebration (toast) & C & Explore and choose a place to celebrate a special occasion with champagne/soft drink \\
Celebration (confetti) & CC & Explore and choose a place to celebrate a special occasion shooting the confetti gun \\ \hline \hline
\end{tabular}
\end{adjustbox}
\end{table}

An open call for participation was issued to the different associations or public centres in the neighbourhood. Three communities agreed to join the project: the school ``Escola Ferrer i Guàrdia (EFiG)'', the neighbourhood association ``Associació de Veïns Sota el Cam\'i Ral (AAVV)'', and the older adults group ``Espai Actiu de la Gent Gran (EAGG)''. Three 2-hour presentation sessions were organized with the three participating communities, hosted by the researchers. The aim of these sessions was to present the project to the interested members of each community, along with all the details corresponding to the objectives and the collection of data on pedestrian mobility. All doubts and questions were discussed. The meetings represented a chance to reflect on the urban structure of the neighbourhood. The attendees discussed the form of the pedestrian mobility experiments, and were actively engaged in the creation of the experimental protocol for the experiments.

The first meeting was held at the ``Associaci\'o de Ve\"ins Sota el Cam\'i Ral (AAVV)'' on February 16. The second meeting took place at the ``Espai Actiu de la Gent Gran (EAGG)'' on March 2. The last meeting was held on March 16 at the ``Escola Ferrer i Guàrdia (EFiG)''. People interested in doing the experiment had about one month and a half to form the groups and confirm their participation in one of the three local communities involved. In this way, we were able to know the number of participating groups in advance and thus had enough time to prepare the material for all the experiments (e.g., the tablet devices and the festive exploration kits, etc.). The research team did not collect any personal data and only knew the number of groups and participants. The call for participation resulted in 19 groups of 3-5 people and a total number of 72 participants. From the school (EFiG), there were 15 teachers who formed four different groups and a classgroup (6th grade of primary school, 11-12 years old) of 26 students, who formed seven groups. From the two associations (AAVV and EAGG), eight groups were formed with a total of 31 participants.

\subsection*{Data acquisition with the tablet device App}

A set of tasks were proposed to the participants, which they had to complete in about 1h 30m. The tasks consisted in performing specific actions of a festive nature, and the groups participating in the experiment had to find their preferred place inside the neighbourhood. The exploration kit included the objects to perform the set of actions and the related instructions which read: ``Find a place to dance with your group. Connect the loudspeaker from the kit via Bluetooth and play a song with the iPad'' or ``Choose a place to make a toast: open the drink and make a toast to a special occasion for you''. The list of the six different tasks is described in Table \ref{tab:missions}. The aim of the kit was to create an enjoyable recreational   experience and to encourage the participants to adopt an exploratory attitude towards the neighbourhood. 

The experiments with the participating groups from the school (EFiG) took place on Friday April 29, 2022, during school hours. The groups started and finished the data collection in the vicinity of the school. The four groups of teachers participated in the experiment during the lunch break (between 1:30 and 3:00 pm). The seven groups of students from the same class participated as part of their timetable (3:30 - 5:00 pm). The following day, Saturday April 30 (11:30 am - 1:00 pm), the eight remaining groups (AAVV and EAGG) participated. These groups started gathering data at the EAGG and finished at ``Pla\c ca de la Font'', a square in the middle of the neighbourhood (see Figure \ref{fig:map_granollers}). The weather conditions were favourable on both days: there were no exceptional climate events (e.g., rain or very low/high temperatures).

Each participating group received a tablet device (iPad). On each of the devices, the Wikiloc App had been previously installed and a user id (Google account) created by the research staff was already logged in. Thus, the participants did not have to provide any personal information to use the App and the researchers were able to download the GPS data collected in {\tt gpx} format once the experiment ended. A single account was assigned to each tablet device. Since there were eight tablets, only eight groups could perform the experiment simultaneously. Few groups used the same tablet device (at different times) and hence the same account. For this reason, a code name was assigned to each participating group. The group used the code name to save the recorded trajectory and the file generated {\tt gpx} already contained the group code in the file name. The protocol thus avoided the possibility that participants might introduce personal data (e.g., their own name) in the file name. Finally, every tablet device was tested days prior to the experiment by the researchers, recording a trajectory with the Wikiloc App and exporting the GPS data.

\subsubsection*{Scientific protocol for the experiment}

\begin{figure}
\centering
\includegraphics[width=0.55\textwidth]{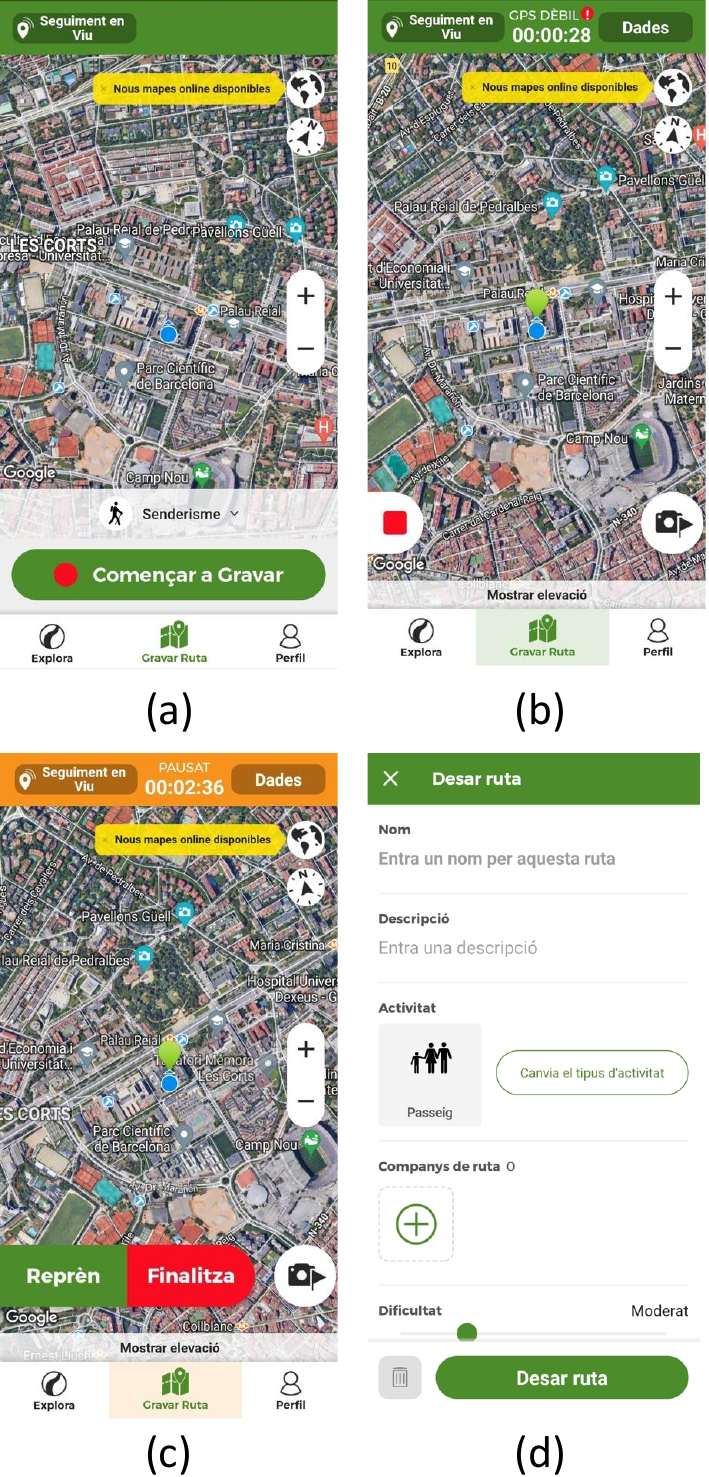}
\caption{{\bf Wikiloc App screenshots (in Catalan) during a test carried out in the University facilities (Faculty of Physics, UB).} (a) Initial screen when opening the Wikiloc App. Participants had to click on ``Comen\c ar a Gravar'' (Start recording, in Catalan) to start recording the trajectory. (b) A real time recording of the journey; the recording time is displayed at the top of the screen, and the recording can be paused by pushing the red lower button. (c) The screen after pausing the recording of the journey; the participants had to click on ``Finalitza'' (End, in Catalan) to end the experiment and save the GPS data. (d) The screen after finalizing the recording. In this step, the participants returned the tablet to one of the researchers, who in turn saved the GPS data by clicking on ``Desar ruta'' (Save route, in Catalan), after entering the trajectory's name ``Nom'' (Name, in Catalan) which was the code of the respective group.}
\label{fig:screenshots}
\end{figure}

At the meeting point (EFiG for teachers and students on April 29 and EAGG on April 30 for the rest of the groups), the tablets were distributed to the different groups. The participants received the following instructions accompanied by some screenshots (see Figure \ref{fig:screenshots}):

{\bf Before the experiment.} (1) Each group member should fill out the anonymous socio-demographic data form (see Table \ref{tab:sociodem}) and sign the informed consent document.  (2) The group must choose one member to be in charge of starting the App at the beginning of the experiment and stopping it at the end, and another member responsible for carrying the tablet with him/her throughout the journey (preferably someone already familiar with the technology). (3) The research team provides a tablet to the group in which the Wikiloc App is open and ready to start recording the GPS locations (see Figure \ref{fig:screenshots}a) jointly with a list of tasks to complete along with the exploration kit (see Table \ref{tab:missions}).

{\bf During the experiment.} (1) Click on ``Comen\c car a gravar'' (Start recording, in Catalan, see Figure \ref{fig:screenshots}a) when you start the journey. Avoid starting the experiment in indoor spaces as the GPS will lose precision. Walk between 50-100 metres before clicking on the App icon to start the experiment. This ensures a good connection to the GPS satellites. On the screen, you should see the recording time at the top and a red icon to pause the recording (see Figure \ref{fig:screenshots}b). The App is collecting the GPS data from your journey. You do not need to have the Wikiloc visible on your screen; you can use other Apps during your journey. (2) During the experiment, the group must walk together and explore the neighbourhood while completing the list of tasks with the exploration kit. The neighbourhood is limited by the streets: ``Carrer d'Enric Prat de la Riba'', ``Carrer de Roger de Flor'', ``Carrer de Pius XII'' and ``Carrer d'Isabel de Villena'' (see Figure \ref{fig:map_granollers}). The groups have 1 hour and 30 minutes to complete the list of tasks in the list. (3) Use the tablet device to record a video of at least 30 seconds while performing the actions that comprise each task. This will allow the researchers to make a more accurate analysis of the locations chosen to perform the different actions.  

{\bf After the experiment.} (1) Upon completion of the tasks, go to the meeting point - ``Pla\c ca de la Font'' for the groups belonging to the associations AAVV and/or EAGG (April 30), and EFiG for the participants from the school (April 29), see Figure \ref{fig:map_granollers}. (2) Press the red icon to pause the recording (see Figure \ref{fig:screenshots}b) and then ``Finalitza'' (End, in Catalan) to end the GPS recording (see Figure \ref{fig:screenshots}c). (3) The App will then direct you to the save screen (see Figure \ref{fig:screenshots}d). (4) Return the tablet to the researchers who are at the meeting point. They will check if you have correctly named and saved the recorded trajectory (using the group code name).

\begin{table}[t]
\centering
\caption{\label{tab:sociodem} {\bf Socio-demographic form in Catalan}. Each participant had to fill out the socio-demographic form, firstly at the top adding the code of the group to which they belong ("Grup") and the number of participants in the group ("Nombre de participants en el grup"). Each participant (each column) in the group recorded their age range ("Rang d'edat"), the gender identity ("Identitat de gènere") male, female, non-binary or I don't want to answer. They also had to answer the question \textit{Do you know the neighbourhood?} ("Conèixes el barri?") with a three-level answer: well, a little, or not at all. Finally, participants had to answer the question \textit{Do you frequently go to the neighbourhood?} ("Freqüentes el barri?) with answers every day, weekly, occasionally or almost never.}
\begin{adjustbox}{width=\textwidth}
\begin{tabular}{|c|c|c|c|c|c|c|c|c|}
\multicolumn{3}{l}{Grup:} \\ 
\multicolumn{3}{l}{Nombre de participants en el grup:} \\ 
\multicolumn{9}{l}{} \\ \hline
\textbf{PARTICIPANTS} & \textbf{P1} & \textbf{P2} & \textbf{P3} & \textbf{P4} & \textbf{P5} & \textbf{P6} & \textbf{P7} & \textbf{P8}  \\ \hline
\multicolumn{9}{|l|}{\textbf{Rang d'edat}} \\ \hline
<17 & & & & & & & &  \\ \hline
18-35 & & & & & & & &  \\ \hline
36-50 & & & & & & & &  \\ \hline
51-65 & & & & & & & &  \\ \hline
>65 & & & & & & & &  \\ \hline
\multicolumn{9}{|l|}{\textbf{Identitat de g\`enere}} \\ \hline
Dona & & & & & & & &  \\ \hline
Home & & & & & & & &  \\ \hline
No binari & & & & & & & &  \\ \hline
No vull respondre & & & & & & & &  \\ \hline
\multicolumn{9}{|l|}{\textbf{Con\`eixes el barri?}} \\ \hline
Molt & & & & & & & &  \\ \hline
Poc & & & & & & & &  \\ \hline
Gens & & & & & & & &  \\ \hline
\multicolumn{9}{|l|}{\textbf{Freq\"uentes el barri?}} \\ \hline
Cada dia & & & & & & & &  \\ \hline
Setmanalment & & & & & & & &  \\ \hline
Ocasionalment & & & & & & & &  \\ \hline
Gens & & & & & & & &  \\
 \hline
\end{tabular}
\end{adjustbox}
\end{table}

\subsubsection*{Group code names}
As mentioned in the protocol, each group received a tablet device with the Wikiloc App with a session opened and ready to start recording the trajectory. A code was assigned to each group, which was also used to fill out the socio-demographic form and to write the session name with the GPS data collected in a single trajectory (and therefore the {\tt gpx} also had the group code name as a file name).

The code names assigned to the four groups of teachers from to the school (EFiG, Friday April 29) were: P1, P2, P3, and P4. The code names assigned to  the students who participated in the experiment in the afternoon (EFiG, Friday April 29) were: E1, E2, E3, E4, E5, E6, and E7. On the following day, April 30, local participants from the residents' associations and older people (AAVV and EAGG) had the eight following codes: G1, G2, G3, G4, G5, G6, G7, and G8. For each group (code name), Table \ref{tab:GPS_data} shows the number of participants, the number of GPS data records and the number of stops, before and after data processing and filtering. Tables \ref{tab:sociodem2}, \ref{tab:sociodem3} and \ref{tab:sociodem4} show the results of the socio-demographic form filled in by each participating group. The profile of the 72 participants was quite heterogeneous in terms of age due to the different profile of each community involved. There was a higher participation of women (44) than men (28) and in most cases the participants reported frequenting the neighbourhood and knowing it well.

\begin{table}[h]
\centering
\caption{\label{tab:GPS_data} \textbf{Number of participants, GPS data records and stops per each participant group}. The columns represent the profile of the participants, the code names for each group, the number of participants and the number of GPS records and stops. We add the original data being collected and the data after filtering and cleaning processes. The last row includes the total number of participants, the GPS records and the number of stops identified. The participating groups are split in three main blocks corresponding to the three experimental sessions held: teachers (EFiG), students (EFiG), and residents jointly with the older adults (AAVV and EAGG). P3 failed in the data recording and saving process.}
\begin{adjustbox}{width=\textwidth}
\begin{tabular}{cccccccc}
\hline \hline
& & & \multicolumn{2}{c}{Original Data} & & \multicolumn{2}{c}{Processed Data} \\ \cline{4-5} \cline{7-8}
& Code name & Participants & GPS records & Stops & & GPS records & Stops \\ \hline
\multirow{4}{4em}{School teachers} & P1 & 3 & 306 & 16 & & 291 & 16 \\
& P2 & 3 & 251 & 19 & & 233 & 17 \\
& P3 & 5 & -- & -- & & -- & -- \\
& P4 & 4 & 175 & 21 & & 150 & 17 \\ \hline
\multirow{7}{4em}{School students} & E1 & 3 & 219 & 25 & & 161 & 18 \\
& E2 & 3 & 159 & 21 & & 133 & 17 \\
& E3 & 4 & 275 & 21 & & 206 & 16 \\
& E4 & 4 & 170 & 27 & & 123 & 20 \\ 
& E5 & 4 & 210 & 32 & & 182 & 29 \\ 
& E6 & 4 & 300 & 27 & & 257 & 21 \\ 
& E7 & 4 & 321 & 45 & & 228 & 32 \\ \hline
\multirow{8}{4em}{Adults, families and older adults} & G1 & 5 & 241 & 40 & & 227 & 36 \\
& G2 & 3 & 72 & 7 & & 52 & 4 \\
& G3 & 3 & 163 & 25 & & 150 & 21 \\
& G4 & 3 & 90 & 10 & & 64 & 5 \\ 
& G5 & 3 & 168 & 26 & & 143 & 20 \\ 
& G6 & 3 & 107 & 26 & & 87 & 13 \\ 
& G7 & 6 & 162 & 25 & & 156 & 22 \\ 
& G8 & 5 & 143 & 17 & & 138 & 15 \\ \hline 
Total & & 72 & 3532 & 420 & & 2981 & 339 \\ \hline \hline
\end{tabular}
\end{adjustbox}
\end{table}

\begin{table}[h]
\centering
\caption{\label{tab:sociodem2} {\bf Socio-demographic statistics of all the participants}. Number of participants according to the socio-demographic form. The columns represent the participating groups in the three experimental sessions that were held (teachers from EFiG, students from EFiG, and people from the neighbourhood jointy with old adults from AAVV and EAGG). The last column  aggregates all participants.}
\begin{adjustbox}{width=\textwidth}
\begin{tabular}{ccccc}
\hline\hline
 & Teachers (EFiG) & Students (EFiG)  & Residents (AAVV/EAGG)  & Total  \\ \hline
 Number of participants & 15 & 26 & 31 & 72 \\ \hline
 \textbf{Gender Identity} &  &  &  &  \\ 
 Woman & 12 & 10 & 22 & 44 \\
 Man & 3 & 16 & 9 & 28 \\ \hline
 \textbf{Age range} & & & & \\
< 17 & 0 & 26 & 7 & 33 \\
17-35 & 1 & 0 & 2 & 3 \\
36-50 & 12 & 0 & 6 & 18 \\
51-65 & 2 & 0 & 8 & 10 \\
>65 & 0 & 0 & 8 & 8 \\ \hline
\textbf{Do you know the neighbourhood?} & & & & \\
Much & 5 & 15 & 21 & 41 \\
Little & 6 & 10 & 7 & 23 \\
None & 4 & 1 & 3 & 8 \\ \hline
\textbf{Do you frequently visit the neighbourhood?} & & & & \\
Every day & 6 & 14 & 22 & 42 \\
Weekly & 2 & 11 & 3 & 16  \\
Rarely & 1 & 0 & 6 & 7 \\
Almost never & 6 & 1 & 0 & 7 \\
 \hline
 \hline
\end{tabular}
\end{adjustbox}
\end{table}

\begin{table}[h]
\centering
\caption{\label{tab:sociodem3} {\bf Socio-demographic statistics of the school groups (EFiG)}. Number of participants according to the socio-demographic form for the school EFiG. There were seven groups of students (26 participants) and four of teachers (15 participants).}
\begin{adjustbox}{width=\textwidth}
\begin{tabular}{ccccccccccccc}
\hline\hline
& \multicolumn{7}{c}{Students EFiG (26)} & & \multicolumn{4}{c}{Teachers EFiG (15)} \\ \cline{2-8} \cline{10-13}
& E1 & E2 & E3 & E4 & E5 & E6 & E7 & & P1 & P2 & P3 & P4 \\ \hline
Number of participants & 3 & 3 & 4 & 4 & 4 & 4 & 4 & & 3 & 3 & 5 & 4 \\ \hline
\textbf{Gender Identity} &  &  &  &  & & & & & & & & \\ 
Woman & 3 & 2 & 1 & 2 & 0 & 1 & 4 & & 3 & 1 & 5 & 3 \\
Man & 0 & 1 & 3 & 2 & 4 & 3 & 0 & & 0 & 2 & 0 & 1 \\ \hline
\textbf{Age range} &  &  & &  &  & & & & & & & \\
< 17 & 3 & 3 & 4 & 4 & 4 & 4 & 4 & & 0 & 0 & 0 & 0 \\
17-35 & 0 & 0 & 0 & 0 & 0 & 0 & 0 & & 0 & 0 & 1 & 0 \\
36-50 & 0 & 0 & 0 & 0 & 0 & 0 & 0 & & 2 & 3 & 4 & 3 \\
51-65 & 0 & 0 & 0 & 0 & 0 & 0 & 0 & & 1 & 0 & 0 & 1 \\
>65 & 0 & 0 & 0 & 0 & 0 & 0 & 0 & & 0 & 0 & 0 & 0 \\ \hline
\textbf{Do you know the neighbourhood?} &  &  &  &  & & & & & & & & \\ 
Much & 1 & 3 & 1 & 2 & 4 & 3 & 1 & & 2 & 1 & 2 & 0 \\
Little & 2 & 0 & 3 & 2 & 0 & 0 & 3 & & 1 & 1 & 0 & 4 \\
None & 0 & 0 & 0 & 0 & 0 & 1 & 0 & & 0 & 1 & 3 & 0 \\ \hline
\textbf{Do you frequently visit the neighbourhood?} &  &  &  & & & & & & & & & \\ 
Every day & 0 & 3 & 1 & 1 & 3 & 3 & 3 & & 1 & 1 & 0 & 4 \\
Weekly & 3 & 0 & 3 & 3 & 1 & 0 & 1 & & 0 & 2 & 0 & 0 \\
Rarely & 0 & 0 & 0 & 0 & 0 & 0 & 0 & & 0 & 0 & 1 & 0 \\
Almost never & 0 & 0 & 0 & 0 & 0 & 1 & 0 & & 2 & 0 & 4 & 0 \\
 \hline
 \hline
\end{tabular}
\end{adjustbox}
\end{table}

\begin{table}[t]
\centering
\caption{\label{tab:sociodem4} {\bf Socio-demographic statistics of participating groups from the community (AAVV/EAGG)}. Number of participants according to the socio-demographic form for each group of the two local communities (AAVV and EAGG). There were eight groups (31 participants).}
\begin{adjustbox}{width=\textwidth}
\begin{tabular}{ccccccccc}
\hline\hline
& \multicolumn{8}{c}{AAVV and EAGG (31)}  \\ \cline{2-9}
& G1 & G2 & G3 & G4 & G5 & G6 & G7 & G8 \\ \hline
Number of participants & 5 & 3 & 3 & 3 & 3 & 3 & 6 & 5 \\ \hline
\textbf{Gender Identity} &  &  &  &  & & & &  \\ 
Woman & 3 & 2 & 1 & 3 & 3 & 2 & 4 & 4 \\
Man & 2 & 1 & 2 & 0 & 0 & 1 & 2 & 1 \\ \hline
\textbf{Age range} &  &  &  &  & & & & \\
< 17 & 0 & 0 & 0 & 0 & 0 & 0 & 4 & 3  \\
17-35 & 0 & 0 & 0 & 0 & 0 & 0 & 1 & 1 \\
36-50 & 0 & 0 & 2 & 0 & 1 & 1 & 1 & 1  \\
51-65 & 3 & 2 & 1 & 1 & 1 & 0 & 0 & 0   \\
>65 & 2 & 1 & 0 & 2 & 1 & 2 & 0 & 0  \\ \hline
\textbf{Do you know the neighbourhood?} &  &  &  &  & & & &  \\ 
Much & 0 & 3 & 3 & 2 & 2 & 0 & 6 & 5  \\
Little & 5 & 0 & 0 & 1 & 1 & 0 & 0 & 0  \\
None & 0 & 0 & 0 & 0 & 0 & 3 & 0 & 0  \\ \hline
\textbf{Do you frequently visit the neighbourhood?} &  &  &  &  & & & &  \\ 
Every day & 0 & 2 & 3 & 3 & 3 & 0 & 6 & 5 \\
Weekly & 0 & 1 & 0 & 0 & 0 & 2 & 0 & 0  \\
Rarely & 5 & 0 & 0 & 0 & 0 & 1 & 0 & 0  \\
Almost never & 0 & 0 & 0 & 0 & 0 & 0 & 0 & 0  \\
 \hline
 \hline
\end{tabular}
\end{adjustbox}
\end{table}

\subsubsection*{Supervision, monitoring and support}

\begin{figure}
\centering
\includegraphics[width=0.99\textwidth]{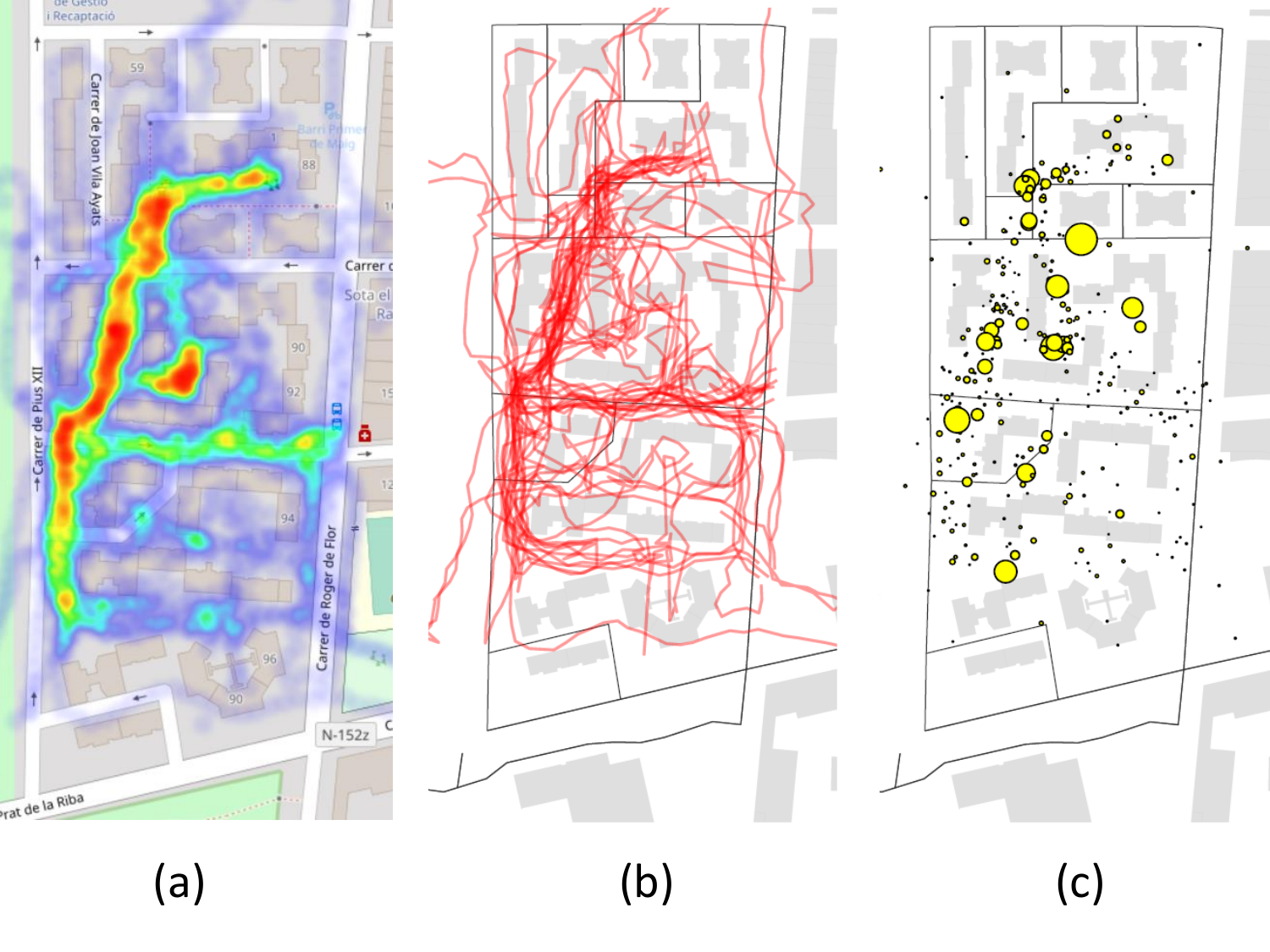}
\caption{{\bf GPS trajectories and stops of all participating groups}. (a) Heat-map of the GPS locations (in motion). (b) Display of all trajectories in the neighbourhood. (c) All the GPS locations labelled as stopped (the size of the yellow circle depends on the duration of the stop: the larger the circle, the longer the stop).  }
\label{fig:maps}
\end{figure}

Before the experiments, scientists from the OpenSystems group tested each tablet, installing the Wikiloc App and recording several trajectories. The {\tt gpx} files with the GPS data were exported and analyzed to ensure that the mobility data were collected accurately and were of good quality. During the experiments, all the research staff also walked through the neighbourhood jointly with the groups to supervise and monitor the progress, and solved any doubts regarding the tasks to be completed or any technical problems with the devices. After the experiment (on the second day, April 30), OpenSystems researchers exported all the {\tt gpx} files with the recorded journeys and created some visualizations of the trajectories on maps to obtain a first impression of the sites the participants chose to perform the tasks and which streets were the most frequented (see Figure \ref{fig:maps}). These preliminary results were shown and discussed with the participating groups during a public lunch that was held at the end of the experiments on the occasion of the neighbourhood festivities.

\subsubsection*{Actionable data to deliver policy recommendations}

One of the main objectives was to work in the neighbourhood as a network. Due to its low cost, its great impact, and the fact that it encourages the residents' active participation, the Design for City Making Research Lab proposed to work on an intensive urbanization strategy (as opposed to the extensive strategies conventionally used in public space urbanization) based on the GPS data collected. The data were carefully analyzed and were combined with a public space analysis that identified the spaces of opportunity for higher impact. Within those spaces, several interventions were proposed in specific locations (see Filtering and processing GPS data Section Filtering and processing GPS data). These interventions were conceived as small constructions including utilities and actionable elements aimed to rehabilitate public spaces. These support structures were meant to socially activate their immediate environment by prompting temporary uses of public space initiated autonomously by citizens. Through this format, the aim was to achieve a large social impact with a cost 10 times lower than that of conventional public space urbanization, while working in conjunction with the municipal departments of urban planning and community action. A report was delivered to the municipality and to the local communities during meetings \cite{Elisava2022}.

\subsection*{Filtering and processing GPS data}

Due to the urban context of the experiment, we were interested in keeping the trajectories of the participating groups within the boundaries of the neighbourhood (Figure \ref{fig:map_granollers}). For this reason, we performed a simple filtering process in which we removed the GPS records at the start and/or end of the trajectory that were outside the neighbourhood. In addition, we performed a more advanced processing in which we labelled each GPS record as stopped or in motion. We also computed the time  difference between consecutive timestamps, the distance, and the instantaneous velocity between consecutive GPS records. We added this extra information in the processed datasets and saved them as new {\tt csv} files. Finally, we also identified the coordinates where the actions took place and the duration of these actions, also saving them in new {\tt csv} files.

\subsubsection*{GPS records beyond the neighbourhood boundaries}

\begin{figure}
\centering
\includegraphics[width=0.9\textwidth]{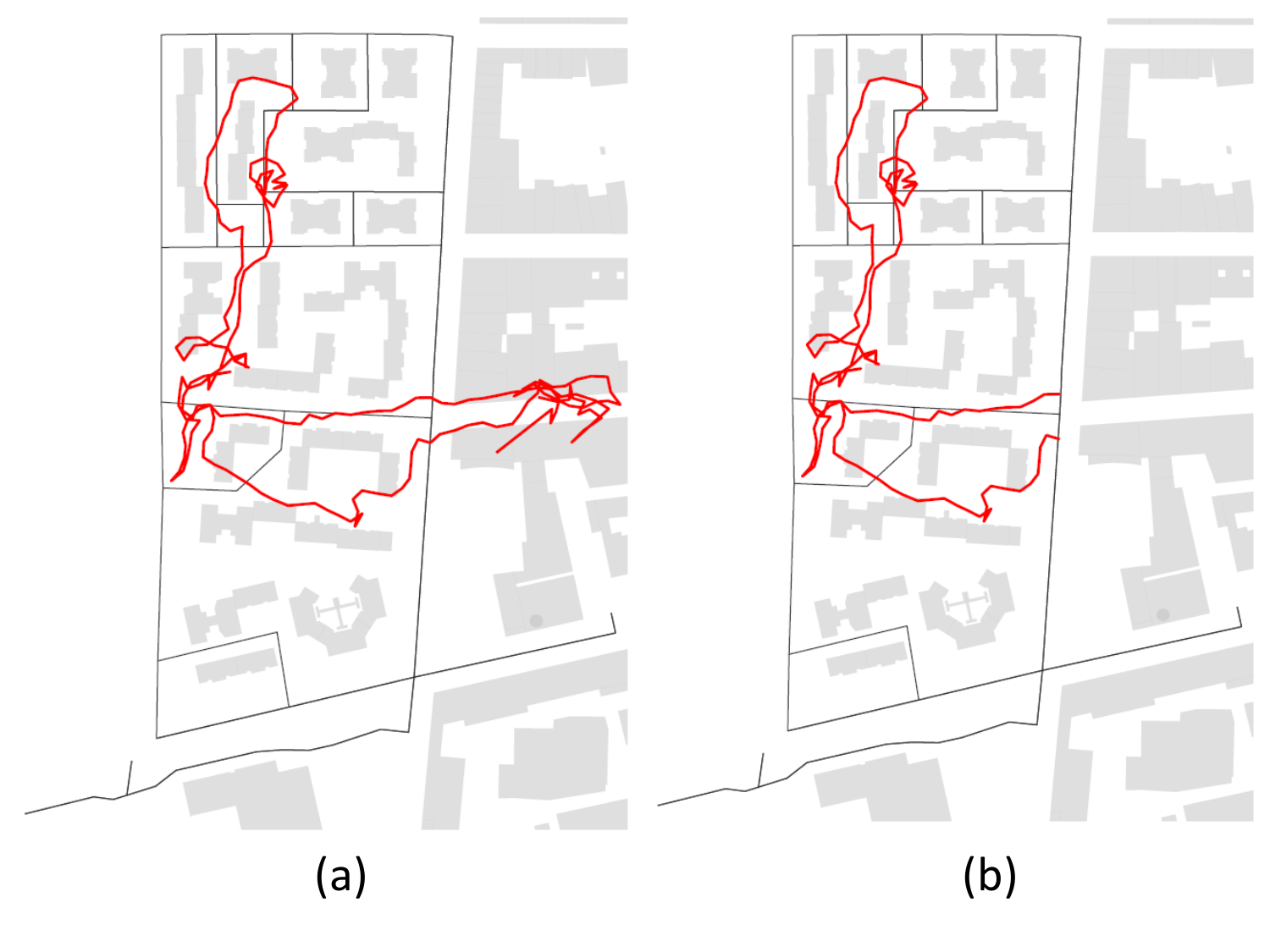}
\caption{{\bf Example of a trajectory clean-up.} The trajectory corresponds to a group from the school EFiG. (a) Map representation of the original GPS location data from a single trajectory. (b) Map representation after the clean-up process, with the removal of the records outside the neighbourhood at the beginning and at the end of experiment.}
\label{fig:process_trajectory}
\end{figure}

During the first seconds of movement (still outside the neighbourhood), GPS records may be noisy due to poor GPS connection (the participants were inside the building) or because the devices were still adjusting the satellites' signal accuracy. However, experiments started slightly outside the established boundaries of the neighbourhood (less than 70 metres, see Figure \ref{fig:map_granollers}) and the manual removal of the records outside the boundaries was enough to avoid noisy GPS data. In addition, the participants from the school (EFiG) also finished outside the boundaries. In this case, it was necessary to remove not just the first data records of the school groups' trajectories but the last part as well (outside the established limits). Figure \ref{fig:process_trajectory}a shows an example of a journey started and finished at the school. The GPS records outside the limits of the district were carefully removed (see Figure \ref{fig:process_trajectory}b). It should be noted that one group of teachers (group P1), went outside the established limits during the experiment, walking through the left side of the neighbourhood (see Figure \ref{fig:map_granollers} and Figure \ref{fig:maps}) where the football field and a green area are located. We decided to keep the full trajectory because they carried out some of the tasks in that space, and also to avoid splitting the trajectory into two different transects. 

In this cleaning process, a total of 551 GPS records were removed from the dataset, thus reducing the final number of GPS records from 3\,532 to 2\,981. Table \ref{tab:GPS_data} shows the number of GPS records and the number of stops before and after this filtering process.

\subsubsection*{Definition of a stop}
Data processing also identified the stops made during the trip. We tagged each GPS record either as stopped or in motion. The Wikiloc App has an automatic pause algorithm, when it detects that there is no movement. We performed reverse engineering to detect the stops:

\begin{enumerate}
\item We first obtain the difference between consecutive GPS timestamps in seconds, $\Delta_{i} (t) = t_{i+1}-t_{i}$. Note that $\Delta_{i} (t$) is computed in a time-advanced way: the time difference of the record at the time stamp $t_{i}$ is the time difference between the timestamps $t_{i}$ and $t_{i+1}$. The duration is irregular and can vary from a few seconds (4, 5, 6...) to long durations, when participants perform the action related to each task. 
\item Then we consider that the device at the timestamp $t_i$ is stopped if $\Delta_{i}(t)$ is greater than or equal to a certain threshold $\Delta_{c}$. After several in-the-field tests, we came up with a threshold of $\Delta_{c}=10$s as the optimal value to capture the stops performed during the trajectory (see section {\tt Technical Validation}). We thus label each GPS record as {\tt stop} or {\tt moving} in the dataset.
\end{enumerate}

Table \ref{tab:GPS_data} shows the number of participants, records (GPS locations) and the number of stops of each group (journey) before and after processing the collected data. The number of detected stops fell by 19.2\% (from 420 to 339) when only the stops within the neighbourhood were considered. The number of stops includes not just stops made to complete the list of actions but also stops due to the actual mobility patterns (crossing the street, changing orientation, etc.). 

\subsubsection*{Distance and instantaneous velocity}
Data processing also includes the distance between geo-locations and the corresponding instantaneous velocity. Both are key variables for characterizing the mobility of each participating group. The distance between consecutive GPS records reads
\begin{equation}
d(t)=|\vec{r}(t+\Delta (t))-\vec{r}(t)|,
\label{d}
\end{equation}
where $\vec{r}(t)$ is the GPS two-coordinate vector of a pedestrian position at time $t$ (a given GPS timestamp) and we remove the label $i$ henceforth for simplicity. We must take into account that the duration of consecutive records $\Delta (t)$ is irregular. We can also define instantaneous velocity as 
\begin{equation}
v(t)=\frac{|\vec{r}(t+\Delta(t))-\vec{r}(t)|}{\Delta(t)}=\frac{d(t)}{\Delta(t)}.
\label{v}
\end{equation}
As for the case of $\Delta (t)$, the calculation of $d(t)$ and $v(t)$ is done in a time-advanced way.

\subsubsection*{Stops due to actions}
We performed a further processing of the mobility data in which we detected the location of the stops to complete the tasks and the duration of the related action. To do so, we used the videos recorded by the participating groups while they were performing the actions. Through the videos they recorded, we obtained and validated the location of the place they chose to perform each action. 
Then, by looking at the dataset and representing the stops and the trajectories in maps, we were able to identify which of the locations in the dataset labelled as stops correspond to the locations extracted from the videos (actions). We finally used the column of the time difference between consecutive records, $\Delta (t)$, to obtain the duration of the stop (action). If we detected two or more consecutive records labelled as stop, they were considered as a single stop. They were then merged as a single record with the location corresponding to the first detected stop and with a stop duration which is the aggregate stop duration.

\begin{figure}
\centering
\includegraphics[width=0.9\textwidth]{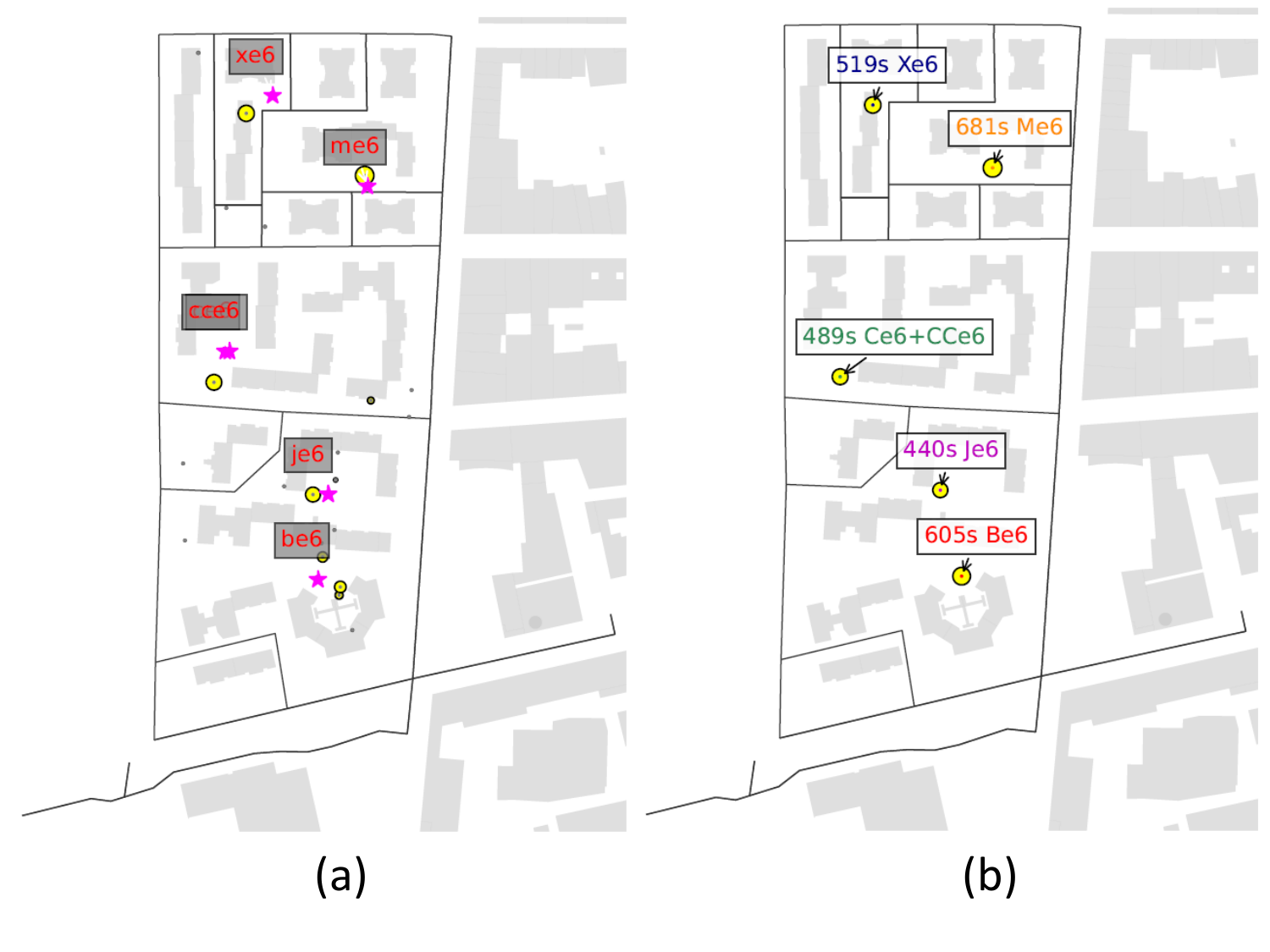}
\caption{{\bf Detecting the stops to complete the tasks and their duration.} Example of a group of students (E6) from the school (EFiG). (a) Shows all the stops detected from the GPS data (yellow dots) together with the locations of the stops to complete the tasks extracted from the recorded videos (purple stars). With the map visualization and the information in the dataset, one can relate each stop location to each task and obtain the duration in seconds. (b) Shows only the stops due to the tasks and their duration in seconds, as well as the code name of each task and group.}
\label{fig:missions_duration}
\end{figure}

Figure \ref{fig:missions_duration} shows an example of the procedure to determine the location of  stops related to each task and its duration. In the first step, the locations of the stops (due to tasks) extracted from the videos are represented in a map (purple stars in Figure \ref{fig:missions_duration}a) with the code of the task and the group. All the labelled stops in the GPS dataset are also represented (yellow dots). One can then easily relate the stops location from the video to the stops detected in the recorded trajectory and obtain the duration of these stops by analyzing the GPS dataset of the group. Occasionally, as stated above, consecutive stops may appear. In these cases, they are considered as a single stop, with their duration the sum of all durations. Then, for clarity, the stops due to the tasks are represented (see Figure \ref{fig:missions_duration}b) indicating the duration and the code of the task and the group code name. A new {\tt csv} file for each group is saved with the information of the location and the duration of the task stops. These new {\tt csv} files do not contain the records of the whole trajectory but only the GPS locations of the stops corresponding to the tasks (and their duration).

\subsection*{Code availability}
The \url{https://github.com/ferranlarroyaub/mobility_granollers} repository contains the Python code and scripts for processing the input data and for replicating the statistical analysis and the figures in this document. The $3.8$ Python version is used to build the code with the main libraries networkx and osmnx to plot the journeys on OpenStreet maps, pandas and numpy to process, clean and analyse the data in Data-Frame format and performing the basic statistical calculations, and scipy for more advanced calculations such as fitting models to the empirical data and matplotlib for plotting purposes. The Python code is built in different Jupyter notebook files which contain the detailed description of the study and the code documentation.

\section*{Data Records}
The data repository \cite{granollersdata} contains both raw and processed datasets, distributed in three different folders. The folder {\tt original data} contains the raw GPS data collected by the participating groups (in {\tt gpx} format). The folder {\tt processed data} contains the same files as the {\tt original data} folder after filtering and cleaning the GPS raw data (see section above) and in {\tt csv} format. Finally, the folder called {\tt stops actions data} contains information about the GPS location and the duration of the actions performed by the different participating groups and related to each task, in {\tt csv} files. We also share the results of the socio-demographic form in the folder called {\tt sociodemographic data}, which contains the {\tt csv} files for each group and for the aggregate of all participants.

\subsubsection*{Raw GPS data}
The {\tt original data} folder contains the raw data collected in the experiment. It consists of 18 different files in {\tt gpx} format, corresponding to each one of the participating groups. One of the groups of teachers (P3), had problems with the tablet device which did not record the trajectory. For this reason there were 18 {\tt gpx} files even though 19 groups participated. Although the trajectory was not recorded, thanks to the video recordings during the tasks in the neighbourhood we have the exact locations where the P3 group stopped to perform the tasks as well as an approximate duration of the stops. The {\tt gpx} file name contains the anonymized nickname (code) of the participant group. The files with the code P ({\tt p1.gxp}, {\tt p2.gpx}, ...) represent the teachers of the school, and those with code E ({\tt e1.gpx}, 
 {\tt e2.gpx}, ...) represent the students of school. Those with code G ({\tt g1.gpx}, {\tt g2.gpx}, ...) represent the people from the neighbourhood and old adults. Each of the {\tt gpx} files has seven columns containing the information described in Table \ref{tab:columns_raw_data}. Table \ref{tab:example_file} shows an example where the number of rows corresponds to the number of records, and so it is different for each participating group.

\begin{table}[t]
\caption{\label{tab:columns_raw_data} {\bf Columns of the raw {\tt gpx} files.} Description of the seven columns of the {\tt gpx} files from the original data, collected using the Wikiloc App.}
\begin{adjustbox}{width=\textwidth}
\begin{tabular}{ll}
\hline\hline
Column & Description \\ \hline
lat & Latitude coordinate of each record in degrees \\ 
lon &  Longitude coordinate of each record in degrees\\ 
name & Code that identifies the group \\ 
cmt & GPS comment of the waypoint \\ 
desc & A text description of the element \\ 
ele & Elevation in meters of the record\\ 
time & Timestamp of each record in  YYYY:MM:DD HH:MM:SS format (UTC) \\ 
 \hline
 \hline
\end{tabular}
\end{adjustbox}
\end{table}

\begin{table}[t]
\caption{\label{tab:example_file} {\bf Example of a raw {\tt gpx} file display.} This is a group of students from the school (EFiG), with 219 records (GPS locations). }
\begin{adjustbox}{width=\textwidth}
\begin{tabular}{cccccccc}
\hline\hline
& lat & lon & name & cmt & desc & ele & time \\ \hline
0& 41.603439 &	2.283413 & E1 & E1 & E1 & 138.518	& 2022-04-29T15:39:40Z \\
1& 41.603425&	2.283494&	E1&	E1	&E1&	139.298	&2022-04-29T15:39:43Z \\
2& 41.603432&	2.283558&	E1&	E1&	E1&	139.917	&2022-04-29T15:39:45Z \\
...&...&...&...&...&...&...&...\\
216&	41.603345&	2.283466&	E1	&E1&	E1&	142.929&2022-04-29T16:35:37Z\\
217&	41.603397&	2.283473&	E1&	E1&	E1	&143.060&	2022-04-29T16:52:34Z\\
218&	41.603241&	2.283207&	E1&	E1&	E1&	143.393&	2022-04-29T16:52:38Z\\
 \hline
 \hline
\end{tabular}
\end{adjustbox}
\end{table}

\subsubsection*{Filtered and processed GPS data}
The {\tt processed data} folder contains the 18 {\tt csv} files reporting the participants' trajectories after the filtering and cleaning process. In the processed data files, the columns {\tt cmt} and {\tt desc} are removed as they do not give relevant information. These columns indicate the comments and the description of the trajectory respectively, and in our dataset they both have the same value as the {\tt name} column, which indicates the name of the trajectory (group code, e.g. E1, see Tables \ref{tab:columns_raw_data} and \ref{tab:example_file}). The column {\tt ele} indicates the elevation (in metres) and is preserved although it was not used in our current related research. Four new columns are also included (see Table \ref{tab:columns_processed_data}), representing the distance, velocity and time increment between consecutive GPS records and the label of each record indicating whether it is stopped (stop) or in motion (moving). The number of columns is then increased to 9. We note again that $\Delta(t)$, $d(t)$ and $v(t)$ are calculated in a time-advanced way (cf. Eqs. (\ref{d}) and (\ref{v})): the time difference of the record at the time stamp $t_{i}$ is the time difference between the timestamps at locations $t_{i}+\Delta_{i}(t)$ and $t_{i}$ (the same applies for the distance and velocity). For this reason, the last record (last row) does not have these three values reported. Table \ref{tab:example_file2} shows an example of the processed data. The processed {\tt csv} files are saved with the same name as the original data (using the code of the group) but adding the suffix {\tt processed} (e.g., {\tt g4\_processed.csv}).

\begin{table}[t]
\caption{\label{tab:columns_processed_data} {\bf New columns of the filtered and processed \textit{csv} files.} Description of the four new columns added to the processed dataset. }
\begin{adjustbox}{width=\textwidth}
\begin{tabular}{ll}
\hline\hline
Column & Description \\ \hline
$\Delta t$ & Time difference between consecutive timestamps $t$, in seconds \\ 
$d$ &  Euclidean distance a time $t$ between consecutive GPS locations, in metres\\ 
$v$ & Instantaneous velocity at time $t$ (distance over time lapse), in metres/second  \\ 
stops & Indicates if the participant is stopped ("stop") or moving ("flight")\\
 \hline
 \hline
\end{tabular}
\end{adjustbox}
\end{table}

\begin{table}[t]
\caption{\label{tab:example_file2} {\bf Example of a processed and cleaned \textit{csv} file display.} This is a participating group from the Association of neighbours and older people (AAVV, EAGG), with $156$ GPS records.}
\begin{adjustbox}{width=\textwidth}
\begin{tabular}{cccccccccc}
\hline\hline
& lat & lon & name & ele & time & $\Delta t$ & d & v & stops\\ \hline
0	&41.602788&	2.282938	&G7	&143.753&	2022-04-30 11:28:20+00:00&	6.0	&6.117981&	1.019664&	moving \\
1&	41.602843&	2.282936&	G7&	143.716&	2022-04-30 11:28:26+00:00&	6.0&	5.306684&	0.884447&	moving \\
2&	41.602889&	2.282919&	G7&	143.832&	2022-04-30 11:28:32+00:00&	7.0&	5.524785&	0.789255&	moving \\
...&...&...&...&...&...&...&...&...&...\\
153&	41.604316&	2.282312&	G7&	154.909&	2022-04-30 13:02:44+00:00&	5.0& 5.823689&	1.164738&	moving\\
154&	41.604325&	2.282243&	G7&	155.159&	2022-04-30 13:02:49+00:00&	4.0	&5.269312&	1.317328&	moving\\
155	&41.604364	&2.282207&	G7&	155.178&	2022-04-30 13:02:53+00:00&	NaN&	NaN&	NaN&	stop\\
 \hline
 \hline
\end{tabular}
\end{adjustbox}
\end{table}

\subsubsection*{Stop duration and the tasks}
The {\tt stops actions data} folder contains 19 {\tt csv}s processed files, corresponding to each of the participating groups. The {\tt csv} files are saved using the code name of the group and the suffix {\tt stops} (e.g., {\tt e6\_stops.csv}). Each file provides detailed information about the stops the groups made to complete the tasks: the geo-location of the stop, the duration and to which task it corresponds (see Table \ref{tab:example_file3} as an example). It is worth mentioning that some groups carried out simultaneous tasks at the same location. This is indicated in the action/task code (e.g., Cp4 + CCp4, see Table \ref{tab:example_file3}). To determine the location of the stops and the action that took place, we used the videos recorded by the participating groups as they carried out each of the tasks around the neighbourhood (see Figure \ref{fig:missions_duration}). The csv-files only include the stops due to the tasks carried out by the participants, that were of major urbanistic relevance. The stops due to the micro-mobility of the participants (due to traffic lights, crossing the street, reorientation processes, etc.) are not included. However, they can be identified with the {\tt stop} column of the processed csv-files with the GPS data records. Table \ref{tab:data_missions_professors}, Table \ref{tab:data_missions_students} and Table \ref{tab:data_missions_local} show the duration of each action carried out by each of the school groups (teachers and students, EFiG) and by each of the local association groups (AAVV and EAGG). 

\begin{table}[t]
\centering
\caption{\label{tab:example_file3} {\bf Example of a processed and cleaned \textit{csv} file display, with the stops due to tasks.} This group comprised teachers from the school (EFiG), with five stops to accomplish the tasks. The column ``accio'' is the code of the action (see Table \ref{tab:missions}) followed by the group code (e.g. p4). The columns ``longitud'' and ``latitud'' represent respectively the longitude and latitude coordinates of the stop and the column ``durada'' is the duration of the stop in seconds.}
\begin{tabular}{ccccc}
\hline\hline
& accio & longitud & latitud & durada\\ \hline
0	&	Cp4+CCp4	&	2.281650	&	41.603890	&	361.0 \\
1		&	Jp4		&2.282149	&	41.603109	&	197.0 \\
2		&	Mp4		&2.281773	&	41.603244	&	1811.0 \\
3	&		Xp4	&	2.282434	&	41.603065	&	295.0 \\
4		&	Bp4		&2.282149	&	41.604199		&258.0 \\
 \hline
 \hline
\end{tabular}
\end{table}

\begin{table}[t]
\centering
\caption{\label{tab:data_missions_professors} \textbf{Duration of the actions/tasks (professors from EFiG).} For each group of the teachers from the school (EFiG), the table contains the list of tasks they completed (with the code, see Table \ref{tab:missions}) and the duration of each task.}
\begin{tabular}{ccc}
\hline \hline
\multicolumn{3}{c}{School Professors}  \\ \hline \hline
Code Group & Task & Duration (s) \\ \hline
 \multirow{4}{4em}{P1} & B & 557 \\
 & J & 2151 \\
 & J & 231 \\
 & X+M & 525 \\ \hline
 \multirow{5}{4em}{P2} & B & 151 \\
 & J & 113 \\
 & X & 200 \\
 & M & 317 \\
 & C+CC & 1240 \\ \hline
 \multirow{4}{4em}{P3} & B & 166 \\
 & J + M & 1060 \\
 & X & 330 \\
 & C+CC & 120 \\ \hline
 \multirow{5}{4em}{P4} & B & 258 \\
 & J & 197 \\
 & X & 295 \\
 & M & 1811 \\
 & C + CC & 361 \\ \hline \hline
\end{tabular}
\end{table}

\begin{table}[t]
\centering
\caption{\label{tab:data_missions_students} \textbf{Duration of the actions/tasks (students from EFiG).} For each group of the students from the school (EFiG), the table contains the list of tasks they completed (with the code, see Table \ref{tab:missions}) and the duration of each task.}
\begin{tabular}{ccc}
\hline \hline
\multicolumn{3}{c}{School Students}  \\ \hline \hline
Code Group & Task & Duration (s) \\ \hline
 \multirow{5}{4em}{E1} & B & 151 \\
 & J & 100 \\
 & J & 249 \\
 & X & 521 \\ 
 & M + C + CC & 985 \\ \hline
 \multirow{3}{4em}{E2} & B & 670 \\
 & J & 342 \\
 & X + M + C + CC & 2345 \\ \hline
 \multirow{5}{4em}{E3} & B & 1130 \\
 & J & 262 \\
 & X & 39 \\
 & M + C & 731 \\ 
 & CC & 1240 \\ \hline
 \multirow{5}{4em}{E4} & B + J & 601 \\
 & B & 87 \\
 & J + X & 403 \\
 & M & 963 \\
 & C + CC & 1090 \\ \hline 
 \multirow{4}{4em}{E5} & B + J & 486 \\
 & J & 117 \\
 & X & 817 \\
 & M + C + CC & 858 \\ \hline
 \multirow{5}{4em}{E6} & B & 605 \\
 & J & 440 \\
 & X & 519 \\
 & M & 681 \\
 & C + CC & 489 \\ \hline
 \multirow{4}{4em}{E7} & B + M & 322 \\
 & B + J & 927 \\
 & X & 492 \\
 & C + CC & 1065 \\ \hline
\end{tabular}
\end{table}

\begin{table}[t]
\centering
\caption{\label{tab:data_missions_local} \textbf{Duration of the actions/tasks (local people from AAVV and EAGG).} For each group of the participants from the associations (AAVV and EAGG), the table contains the list of tasks they completed (with the code, see Table \ref{tab:missions}) and the duration of each task.}

\begin{tabular}{ccc}
\hline \hline
\multicolumn{3}{c}{Local People}  \\ \hline \hline
Code Group & Action & Duration (s) \\ \hline
 \multirow{6}{4em}{G1} & B + J & 909 \\
 & B & 100 \\
 & J & 479 \\
 & J & 1882 \\ 
 & X & 1896 \\
 & M + C & 809 \\ \hline
 \multirow{6}{4em}{G2} & B & 427 \\
 & J & 400 \\
 & X & 360 \\
 & M & 946 \\
 & C & 677 \\
 & CC & 82 \\ \hline
 \multirow{4}{4em}{G3} & J + M & 1845 \\
 & J & 512 \\
 & X & 96 \\
 & C + CC & 2669 \\ \hline
 \multirow{2}{4em}{G4} & X + M & 1243 \\
 & C + CC & 477 \\ \hline 
 \multirow{5}{4em}{G5} & B + J & 690 \\
 & J & 267 \\
 & X + M & 1071 \\
 & C & 1223 \\
 & CC & 277 \\ \hline
 \multirow{3}{4em}{G6} & B & 828 \\
 & J + M & 2677 \\
 & X + C & 795 \\ \hline
 \multirow{2}{4em}{G7} & B + J + X + M + C & 3130 \\
 & CC & 1955 \\ \hline
 \multirow{4}{4em}{G8} & B & 918 \\
 & J & 713 \\
 & X & 400 \\
 & M + C + CC & 656 \\ \hline
\end{tabular}
\end{table}

\subsubsection*{Socio-demographic data}
The data repository \cite{granollersdata} also contains the answers to the socio-demographic form that each participating group filled in. In the folder {\tt sociodemographic data} there are 20 {\tt csv} files, 19 of which correspond to each of the groups. The file name contains the group code (e.g., {\tt g5\_sociodem.csv}). The last {\tt csv} file corresponds to the sociodemographic data of the aggregate of all participants ({\tt all\_sociodem.csv}). Table \ref{tab:example_file_sociodem} shows an example of a csv-file visualization of the socio-demographic form (group G7). Tables \ref{tab:sociodem2}, \ref{tab:sociodem3} and \ref{tab:sociodem4} show the socio-demographic statistics of each participating group and for the aggregate of all participants.

\begin{table}[t]
\caption{\label{tab:example_file_sociodem} {\bf Example of a socio-demographic {\tt csv} file display.} Answers to the socio-demographic form of a group of 6 participants from the AAVV/EAGG communities. The {\tt csv} file displays the code of the group, the community to which it belongs, the socio-demographic question, the possible answers and the number of participants who answered. }
\begin{adjustbox}{width=\textwidth}
\begin{tabular}{cccccc}
\hline\hline
& group & community & question & answer & number\_of\_participants \\ \hline
0 & g7 & AAVV/EAGG & Gender Identity & Woman & 4  \\
1 & g7 & AAVV/EAGG & Gender Identity & Man & 2 \\
2 & g7 & AAVV/EAGG & Gender Identity & Non-binary & 0 \\
...&...&...&...&...&...\\
4 & g7 & AAVV/EAGG & Age range & <17 & 4  \\
5 & g7 & AAVV/EAGG & Age range & 18-35 & 1 \\
...&...&...&...&...&...\\
14 & g7 & AAVV/EAGG & Do you frequently visit the neighbourhood? & Rarely & 0  \\
15 & g7 & AAVV/EAGG & Do you frequently visit the neighbourhood? & Almost never & 0 \\
 \hline
 \hline
\end{tabular}
\end{adjustbox}
\end{table}

\section*{Technical Validation}
In order to define the threshold $\Delta_{c}$, we performed several in-the-field tests, in which we recorded the trajectory with the Wikiloc App and made some predetermined stops in specific places. Then we exported the {\tt gpx} data and we used the definition of stop described in the section {\tt Methods} to detect the stops along the journey by taking different values for the threshold $\Delta_{c}=7,8,9,10,...$ seconds. If we use a notably small $\Delta_{c}$, we expect to overestimate the number of stops, thus erroneously detecting extra stops. Conversely, if $\Delta_{c}$ is too large, we will underestimate the number of stops, missing the detection of possible stops on the trajectory. Therefore, we choose the minimum $\Delta_{c}$ such that the method captures the stops performed (and which we know in advance, before performing the tests). 

\begin{figure}
\centering
\includegraphics[width=0.9\textwidth]{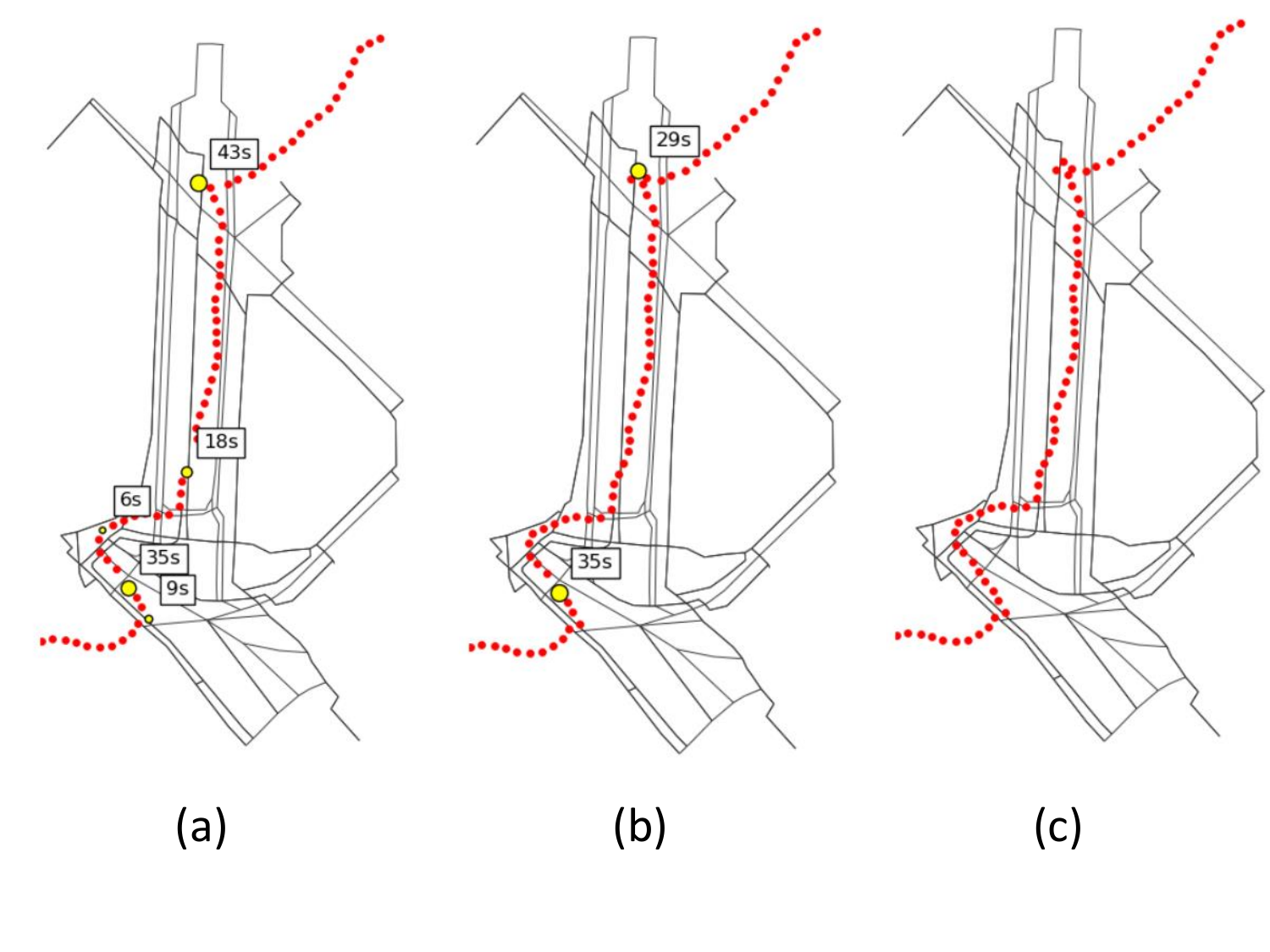}
\caption{{\bf Example of the method to capture the stops in a trajectory.} Various thresholds $\Delta_{c}$ are applied to detect which is the minimum time difference to capture a stop. The red dots represent the track and the yellow dots the stops captured by the method proposed. The duration in seconds is also included. In this particular journey, two stops were made due to traffic lights, of about 30 seconds. (a) Corresponds to a threshold of $\Delta_{c}=6$s, which overestimates the number of stops. (b) Corresponds to the optimal threshold, $\Delta_{c}=10$s, which is the lowest one that captures the two stops. (c) Corresponds to a threshold of $\Delta_{c}=50$s, which underestimates the number of stops, without detecting the two stops. }
\label{fig:stops}
\end{figure}

Consider the example in Figure \ref{fig:stops}. We walked for about six minutes through the ``Esquerra de l'Eixample'' neighbourhood in Barcelona. We made only two stops at two traffic lights of about 30 seconds. In the left panel (Figure \ref{fig:stops}a), a threshold of six seconds is used to detect the stops. As we see, more than two stops are detected. In the right panel (Figure \ref{fig:stops}c), a threshold of 50 seconds is applied, which does not capture any stop. The minimum time threshold that captures the two stops performed is $\Delta_{c}=10$s (middle panel, Figure \ref{fig:stops}b). Other tests were performed, in different places and using different devices, which yielded a threshold of $\Delta_c=10$s as the optimal threshold to capture the stops made during the trajectory. 

\section*{Usage Notes}
The data collected from the experiment represent a rich source of exploratory mobility information for the study of people's micro-behaviour around the neighbourhood ``Primer de Maig'' in the city of Granollers. The mobility data collected in the experiment can be accessed in the folder {\tt original data} from the data repository. Each {\tt gpx} file represents a participant group and their trajectory record. Each individual trajectory belongs to a different group with different socio-demographic characteristics. The Python notebook called {\tt Data Processing and Cleaning.ipynb} from the GitHub repository contains a description of the treatment and data clean-up discussed in this Data Descriptor, as well as the scripts to reproduce all the trajectory representation on maps (Figures \ref{fig:maps}, \ref{fig:process_trajectory} and \ref{fig:stops}). Each processed individual trajectory is saved with the group name (e.g., {\tt p2\_processed.csv}) in a new {\tt csv} file contained in the {\tt processed data} folder. Therefore, clean pedestrian mobility data can also be accessed through this folder. In addition, the Python Notebook mentioned above contains the script for the calculation of the time increment, distance and velocity between consecutive records, as well as the calculation of the stops. These variables are added as new columns to the processed {\tt csv} file of each participant group. The code repository also contains different Jupyter notebooks with the necessary functions and scripts to study and characterize the participants' movement through some statistical patterns which are not discussed in this Data Descriptor. All these files contain the detailed description of the study and the explanation of the code. Finally, through the {\tt stops actions data} folder one can access the locations and the duration of the stops made by each of the participating groups to complete the list of tasks/actions, and through the {\tt sociodemographic data} folder one can access the socio-demographic statistical data.

\section*{Acknowledgements}

We thank the 72 volunteers from the school ``Escola Ferrer i Gu\`ardia (EFiG)'', the associations ``Associaci\'o de Ve\"ins Sota el Cami Ral'' and ``Espai Actiu de la Gent Gran'' for their participation. The study was partially supported by the Ministerio de Ciencia e Innovaci\'on and Agencia Estatal de Investigaci\'on MCIN/AEI/10.13039/501100011033, grant number PID2019-106811GB-C33 (FL and JP); by MCIN/AEI/10.13039/501100011033 and by ``ERDF A way of making Europe", grant number PID2022-140757NB-I00 (FL and JP). We also acknowledge the support of Generalitat de Catalunya through Complexity Lab Barcelona, grant number 2021SGR00856 (FL, JP); and by the ICI programme (Interculturalitat i Cohesi\'o Social) funded by Fundaci\'o Banc\`aria “la Caixa”.

\section*{Authors' contributions}

Ferran Larroya - data acquisition - project conception - data validation – writing - proofreading. Josep Perell\'o - data acquisition - project conception - writing - proofreading. Roger Paez - data acquisition - project conception - writing - proofreading. Manuela Valtchanova - data acquisition - project conception - writing - proofreading.

\section*{Competing interests}

The authors declare that they have no known competing financial interests or personal relationships that could have influenced the work reported in this paper.


\begin{thebibliography}{1}
\expandafter\ifx\csname url\endcsname\relax
  \def\url#1{\texttt{#1}}\fi
\expandafter\ifx\csname urlprefix\endcsname\relax\def\urlprefix{URL }\fi
\providecommand{\bibinfo}[2]{#2}
\providecommand{\eprint}[2][]{\url{#2}}

\bibitem{Ravazzoli2017} Ravazzoli, E., \& Torricelli, G. P. Urban mobility and public space. A challenge for the sustainable liveable city of the future. \emph{The Journal of Public Space} \textbf{2}(2), 37-50 (2017).

\bibitem{Carmona2018} Carmona, M. Principles for public space design, planning to do better. \emph{Urban Des Int} \textbf{24}, 47–59 (2019).

\bibitem{jiang} Jiang, S. et al. The TimeGeo modeling framework for urban mobility without travel surveys. \emph{Proceedings of the National Academy of Sciences} \textbf{113}, E5370–E5378 (2016).

\bibitem{Batty2013} Batty, M. \emph{The new science of cities} (MIT Press, Cambridge, 2013).

\bibitem{Bibri2021} Bibri, S.E. \& Krogstie, J. A Novel Model for Data-Driven Smart Sustainable Cities of the Future: A Strategic Roadmap to Transformational Change in the Era of Big Data. \emph{Future Cities and Environment} \textbf{7}(1): 3, 1–25 (2021).

\bibitem{Bettencour2021} Bettencourt. \emph{Introduction to Urban Science: Evidence and Theory of Cities as Complex Systems} (MIT Press, Cambridge, 2021). 

\bibitem{Bai2016} Bai, X. et al. Defining and advancing a systems approach for sustainable cities. \emph{Current opinion in environmental sustainability} \textbf{23}, 69-78 (2016).

\bibitem{clemente} Xu, Y., Clemente, R. D. \& González, M. C. Understanding vehicular routing behavior with location-based service data. \emph{EPJ Data Sci.} \textbf{10}, 12 (2021). 

\bibitem{hunter} Hunter, R. F. et al. Effect of COVID-19 response policies on walking behavior in US cities. \emph{Nat. Commun.} \textbf{12}, 3652 (2021).

\bibitem{emoro} Yang, Y., Pentland, A. \& Moro, E. Identifying latent activity behaviors and lifestyles using mobility data to describe urban dynamics. \emph{EPJ Data Sci.} \textbf{12}, 15 (2023).

\bibitem{Tobin2022} Tobin, M., Hajna, S., Orychock, K. et al. Rethinking walkability and developing a conceptual definition of active living environments to guide research and practice. \emph{BMC Public Health} \textbf{22}, 450 (2022).

\bibitem{Thomas2016} Thomas, D. \emph{Placemaking: An urban design methodology} (Routledge, New York, 2016).

\bibitem{Paez2019} Paez, R. \emph{Operative Mapping: Maps as Design Tools} (ACTAR publishers, Barcelona, 2019).

\bibitem{Valtchanova2023} Valtchanova, M., \& Paez, R. \emph{Operative Mapping and Collaborative Actions as Design Tools for Critical Socio-spatial Urban Interventions} (Repurposing Places, 133. Proceedings of the International Conference
Repurposing Places for Social and Environmental Resilience
Editor: Anastasia Karandinou
Published 2023)

\bibitem{paez2024} 
Paez, R., Valtchanova, M., Larroya, F. \& Perell\'o, J. \emph{Maps as Design Tools: Space, Time, and Experience. The Routledge Handbook of Cartographic Humanities} (Eds. Tania Rossetto and Laura Lo Presti) 172-181 (Abingdon-on-Thames/New York: Routledge, 2024). https://doi.org/10.4324/9781003327578-22

\bibitem{citizenscience} Vohland, K., Land-Zandstra, A., Ceccaroni, L., Lemmens, R., Perell\'o, J., Ponti, M., Samson, R., \& Wagenknecht, K. \emph{The Science of Citizen Science} (Springer, Cham, 2021).

\bibitem{citizenscience2} Irwin, A. No PhDs needed: how citizen science is transforming research. \emph{Nature} \textbf{562}(7726), pp.480--483 (2018).

\bibitem{citizenscience3} Kapenekakis, I., Chorianopoulos, K. Citizen science for pedestrian cartography: collection and moderation of walkable routes in cities through mobile gamification. \emph{Hum. Cent. Comput. Inf. Sci.} \textbf{7}, 10 (2017).

\bibitem{Palmer2024} Palmer, L. New inroads on community-centric placemaking. \emph{Nat. Cities} \textbf{1}, 2–4 (2024).

\bibitem{Ballerini2021} Ballerini, L., Bergh, S.I. Using citizen science data to monitor the Sustainable Development Goals: a bottom-up analysis. \emph{Sustain Sci.} \textbf{16}, 1945–1962 (2021).

\bibitem{Christine2021} Christine, D. I., \& Thinyane, M. Citizen science as a data-based practice: A consideration of data justice. \emph{Patterns} \textbf{2}(4) (2021).

\bibitem{Sandercock2023} Sandercock, L. \emph{Mapping possibility: Finding purpose and hope in community planning} (Routledge, London, 2023).

\bibitem{Albert2021} Albert, A., Balázs, B., Butkevičienė, E., Mayer, K., Perelló, J. (2021). \emph{Citizen Social Science: New and Established Approaches to Participation in Social Research. In: Vohland, K., et al. The Science of Citizen Science Ch. 7} (Springer, Cham, 2021).

\bibitem{Bonhoure2023} Bonhoure, I., Cigarini, A., Vicens, J. et al. Reformulating computational social science with citizen social science: the case of a community-based mental health care research. \emph{Humanit Soc Sci Commun} \textbf{10}, 81 (2023).

\bibitem{Elisava2022} Paez, R., Valtchanova, M., Perelló, J., Larroya, F. \& Sànchez, E. \emph{Civic Placemaking 3: Disseny, espai públic i cohesió social} (Elisava, 2022).
\url{https://doi.org/10.46467/ElisavaResearch_CivicPlacemaking3}

\bibitem{Sagarra2016} Sagarra, O., Guti\'errez-Roig, M., Bonhoure, I., Perell\'o J. Citizen Science Practices for Computational Social Science Research: The Conceptualization of Pop-Up Experiments. \emph{Front. Phys.} \textbf{3}: 93 (2016).

\bibitem{Perello2022} Perell\'o, J. New knowledge environments: On the possibility of a citizen social science. \emph{Metode Science Studies Journal} \textbf{12}, 25-31 (2022).

\bibitem{Perello2023} Perell\'o, J., Larroya, F., Bonhoure, I., Peter, F. Citizen science for social physics: digital tools and participation. \emph{Eur. Phys. J. Plus} \textbf{139}(7): 572 (2024).

\bibitem{Gutierrez2016} Gutiérrez-Roig, M., Sagarra, O., Oltra, A., Palmer, J.R.B., Bartumeus, F., Diaz-Guilera, A. \& Perell\'o, J. Active and reactive behaviour in human mobility: the influence of attraction points on pedestrians. \emph{Royal Society Open Science} \textbf{3}, 160177 (2016).

\bibitem{Larroya2023} Larroya, F., D\'iaz, O., Sagarra, O. et al. Home-to-school pedestrian mobility GPS data from a citizen science experiment in the Barcelona area. \emph{Sci Data} \textbf{10}, 428 (2023).

\bibitem{kanonymity} Wang, J., \& Kwan, M. P. Daily activity locations k-anonymity for the evaluation of disclosure risk of individual GPS datasets. \emph{International Journal of Health Geographics} \textbf{19}(1), 1-14 (2020).

\bibitem{Sennett2018} Sennett, R. \textit{Building and Dwelling: Ethics for the City} (Farrar, Straus \& Giroux, New York, 2018).

\bibitem{granollersdata} Larroya, F. \& Perelló, J. Explorative pedestrian mobility GPS data from a citizen science experiment in a neighbourhood. \emph{CORA. Repositori de Dades de Recerca} \url{https://doi.org/10.34810/data898} (2023).

\bibitem{barbosa} Barbosa, H. et al. Human mobility: Models and applications. \emph{Phys. Rep.} \textbf{734}, 1–74 (2018).

\end{thebibliography}
\end{document}